\documentclass[12pt,a4paper,twocolumn]{article}
\pdfoutput=1

\usepackage{mathptmx}
\usepackage{graphicx}
\usepackage{times}
\usepackage[utf8]{inputenc}
\usepackage[T1]{fontenc}
\usepackage{multirow}
\usepackage{personal}

\usepackage{ifpdf}
\usepackage[totalwidth=174mm, totalheight=251mm]{geometry}
\usepackage{sectsty}
\sectionfont{\large}
\subsectionfont{\normalsize}
\paragraphfont{\bf}
\usepackage[margin=6pt,font=small,labelfont=bf]{caption}
\usepackage[compact]{titlesec}
\usepackage{natbib}
\ifpdf
 \usepackage{hyperref}
 \usepackage[all]{hypcap}
\fi

\hypersetup{pdfauthor={D\v{z}enan Zuki\'c},pdftitle={A Neural Network Classifier of Volume Datasets}}

\title{\fontsize{18}{18}\selectfont \bf A Neural Network Classifier of Volume Datasets}

\date{} 

\author
{
D\v{z}enan Zuki\'c \hspace{10mm} Christof Rezk-Salama \hspace{10mm} Andreas Kolb
\vspace{5mm}
\\Computer Graphics and Multimedia Systems Group
\\University of Siegen, Germany
\vspace{5mm}
\\H\"olderlinstrasse 3, 57076 Siegen, Germany
\\ \{zukic, rezk, kolb\}@fb12.uni-siegen.de
}

\begin{document}

\abstract
{
Many state-of-the art visualization techniques must be tailored to the specific type of dataset, its modality (CT, MRI, etc.), the recorded object or anatomical region (head, spine, abdomen, etc.) and other parameters related to the data acquisition process. While parts of the information (imaging modality and acquisition sequence) may be obtained from the meta-data stored with the volume scan, there is important information which is not stored explicitly (anatomical region, tracing compound). Also, meta-data might be incomplete, inappropriate or simply missing.

This paper presents a novel and simple method of determining the type of dataset from previously defined categories. 2D histograms based on intensity and gradient magnitude of datasets are used as input to a neural network, which classifies it into one of several categories it was trained with. The proposed method is an important building block for visualization systems to be used autonomously by non-experts. The method has been tested on 80 datasets, divided into 3 classes and a ``rest'' class.

A significant result is the ability of the system to classify datasets into a specific class after being trained with only one dataset of that class. Other advantages of the method are its easy implementation and its high computational performance.


%
}

\begin{titlepage}
\maketitle
\thispagestyle{empty}
\end{titlepage}
\setcounter{page}{2}

\section{Introduction}

Volume visualization techniques have seen a tremendous evolution within
the past years. Nevertheless, the users of volume visualization systems, which are mainly physicians or other domain scientists with only marginal knowledge about the technical aspects of volume rendering, still report problems with respect to usability. The overall aim of current research in the field of volume visualization is to build an interactive rendering system which can be used autonomously by non-experts.

Recent advances in the field of user interfaces for volume visualization, such
as~\cite{Rautek-2007-SLI} and \cite{Rezk06} have shown that semantic models may be tailored to the specific visualization process and the type of data in order to meet these requirements. The semantic information is built upon a priori knowledge about the important structures contained in the dataset to be visualized. A flexible visualization system must thus contain a high number of different semantic models for the huge variety of different examination procedures.

An important building block for an effective volume rendering framework is a classification technique which detects the type of dataset in use and automatically applies a specific semantic model or visualization technique. For example, some methods are created specifically for visualizing MRI scans of the spine or CT scans of the head, and those methods rely on the actual dataset being of that type (i.e. its modality and its anatomical region).

The prior knowledge required for selecting an appropriate visualization technique includes imaging modality, acquisition sequence, anatomical region, as well as other parameters such as chemical tracing compound. That is beyond the information stored in the file system or the meta-data, therefore we  propose a technique which classifies the datasets using a neural network which operates on statistical information, i.e. on histograms of the 3D data itself.

The remainder of the paper is structured as follows: In the next section we review related work important to our paper. Section~\ref{Sec:Classification} describes our proposed method for automatic classification of 3D datasets. In Section~\ref{Sec:Testing} we describe the test environment our solution was integrated in. Section~\ref{Sec:Results} presents and discusses the results of our approach and Section~\ref{Sec:Conclusion} concludes the paper.

For unfamiliar readers, a nice (and relatively short) introduction to feed-forward neural networks is presented by Svozil et al.~\cite{intronn}.

\section{Related work}

The 2D histogram based on intensity and gradient magnitude was introduced
in a seminal paper by Kindlmann and Durkin~\cite{Kindlmann_TransferFunctions}, and extended to
multi-dimensional transfer functions by Kniss et al.~\cite{kniss-vis01}.
Lundstr\"om et al. ~\cite{Lundstrom_LocalHist} introduced local histograms, which utilize a
priori knowledge about spatial relationships to automatically differentiate between
different tissue types. {\v S}ereda et al.~\cite{SeredaBoundaries} introduced the LH histogram to
classify material boundaries.


Tzeng~et~al.~\cite{vis03-TFNeuralNetwork} suggest an interactive visualization system
which allows the user to mark regions of interest by roughly painting the
boundaries on a few slice images. During painting, the marked
regions are used to train a neural network for multi-dimensional classification. Del Rio
et al. adapt this approach to specify transfer functions in an
augmented reality environment for medical
applications~\cite{IPT-EGVE2005:113-120:2005}.
Zhang et al.~\cite{MRI:Zhang} apply general regression neural networks to classify each point of a dataset into a certain class. This information is later used for assigning optical properties (e.g. color).
Cerquides et al.~\cite{Cerquides06} use different methods to classify each point of a dataset. They use this classification information later to assign optical properties to voxels.
While these approaches utilize neural networks to assign optical properties, the method presented here aims at classifying
datasets into categories. The category information is subsequently used as an \emph{a priori} knowledge to visualize the dataset.

Liu et al.~\cite{imageret} classify CT scans of the brain into
pathological classes (normal, blood, stroke) using a method firmly rooted in
Bayes decision theory.

Serlie et al.~\cite{matfrac} also describe a 3D classification method, but their work is focused on
material fractions, not on the whole dataset. They fit the arch model to the LH histogram, parameterizing a single arch function by expected
pure material intensities at opposite sides of the edge (L,H) and a scale
parameter. As a peak in the LH-histogram represents one type of transition,
the cluster membership is used to classify edge voxels as transition types.

Ankerst et al.~\cite{Ankerst993dshape} conduct classification by using a quadratic form
distance functions on a special type of histogram (shell and sector model) of
the physical shape of the objects.


\section{Automatic Classification of Volume Datasets}\label{Sec:Classification}

The method described in this paper was mostly inspired by \cite{nnpos}. In \cite{nnpos}, neural networks are used to position ``primitives'' on the 2D histogram in order to create transfer function aiming at an effective volume visualization. The method presented here is similar in the sense that it uses 2D histograms as inputs to neural networks.

\begin{figure}[htb]
  \scriptsize
  \begin{center}
    \begin{tabular}{ccc}
      \includegraphics{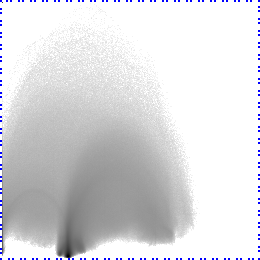} & \includegraphics{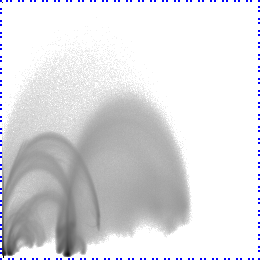} & \includegraphics{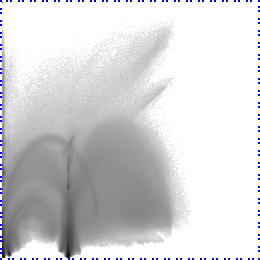}\\
      CTA\_12 & CTA\_19 & CTA\_Sinus\_07\\
      { } & { } & { }\\
      \includegraphics{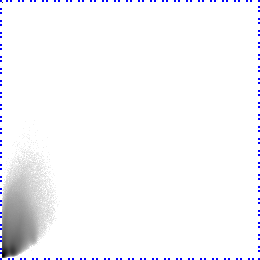} & \includegraphics{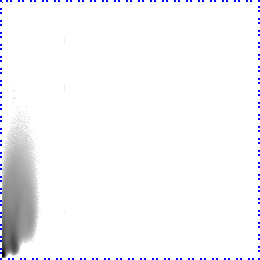} & \includegraphics{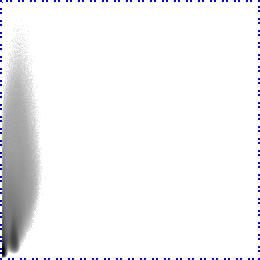}\\
      MR\_02\_interop\_B & MR\_06\_preop & MR\_03\_interop\\
      { } & { } & { }\\
      \includegraphics{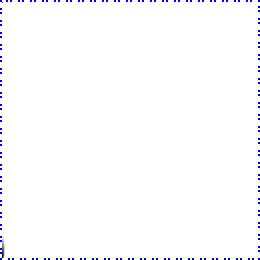} & \includegraphics{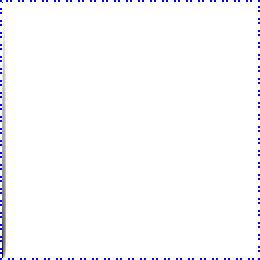} & \includegraphics{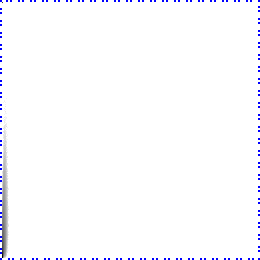}\\
      mr\_ciss\_2 & mr\_ciss\_12 & mr\_ciss\_3\_4\\
      { } & { } & { }\\
      \includegraphics{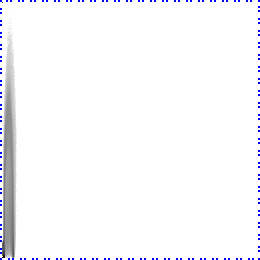} & \includegraphics{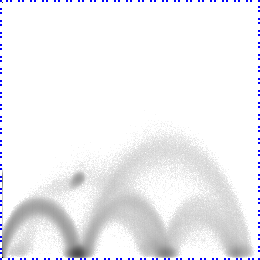} & \includegraphics{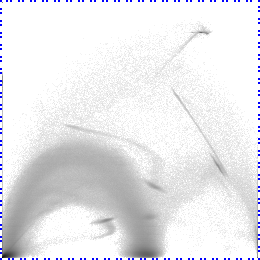}\\
      SpottedHyena256 & tooth\_16 & Engine\\
      { } & { } & { }\\
      \includegraphics{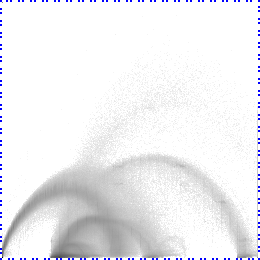} & \includegraphics{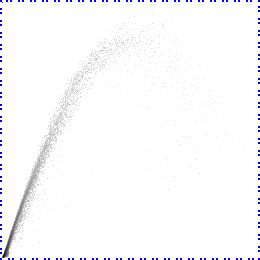} & \includegraphics{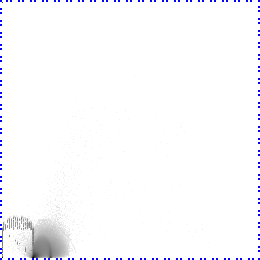}\\
      Tentacle\_combines & Bucky & Woodpecker256
    \end{tabular}
  \end{center}
  \caption{\label{fig:histograms}Some of the histograms. Each one of the first 3 rows represents one class. Histograms in the last two rows together represent the rest class.}
\end{figure}

One of the widely used visualization approaches of 3D data today is direct volume rendering~\cite{rtvg} by means of a 2D transfer function. 2D transfer functions are created in respect to the combined intensity/derivative histogram. Such histograms in turn may be viewed as grayscale images.
All histograms of the same 3D dataset type (like different CT scans of the thorax) look similar to human observers. Likewise, histograms of different datasets types usually look noticeably different (see Fig.~\ref{fig:histograms}). Our method stems from this fact.

Neural networks can easily be trained to approximate many unknown functions for which we have observations in the form of input-output combinations. That makes neural networks suitable for classifying input histograms into categories.

The straight-forward approach is to use the histogram pixels (normalized to the
$[0,1]$ range) as inputs to the neural network. On the output side, each
output corresponds to one category. We take the outputs as representing the probability of the input to belong to the
corresponding category. Thus we have a $k$-dimensional output for $k$
categories. For example, assume that we have the following $[0,1]$ normalized
\footnote{The activation function which is employed in the neural network we used produces outputs
  in the convenient range $[0,1]$, so no additional normalization is necessary}
outputs for some input:
$$\left(
	\begin{array}{c}
	0,893456\\
	0,131899\\
	0,044582
	\end{array}
\right)$$
we interpret them as the probabilities of the input belonging to respective
category (category one -- 89\%, category two -- 13\% and category three --
4\%). Notice that the actual outputs in general do not add up to 100\%.

\begin{figure*}[htbp]
  \centering
  \includegraphics[width=174mm]{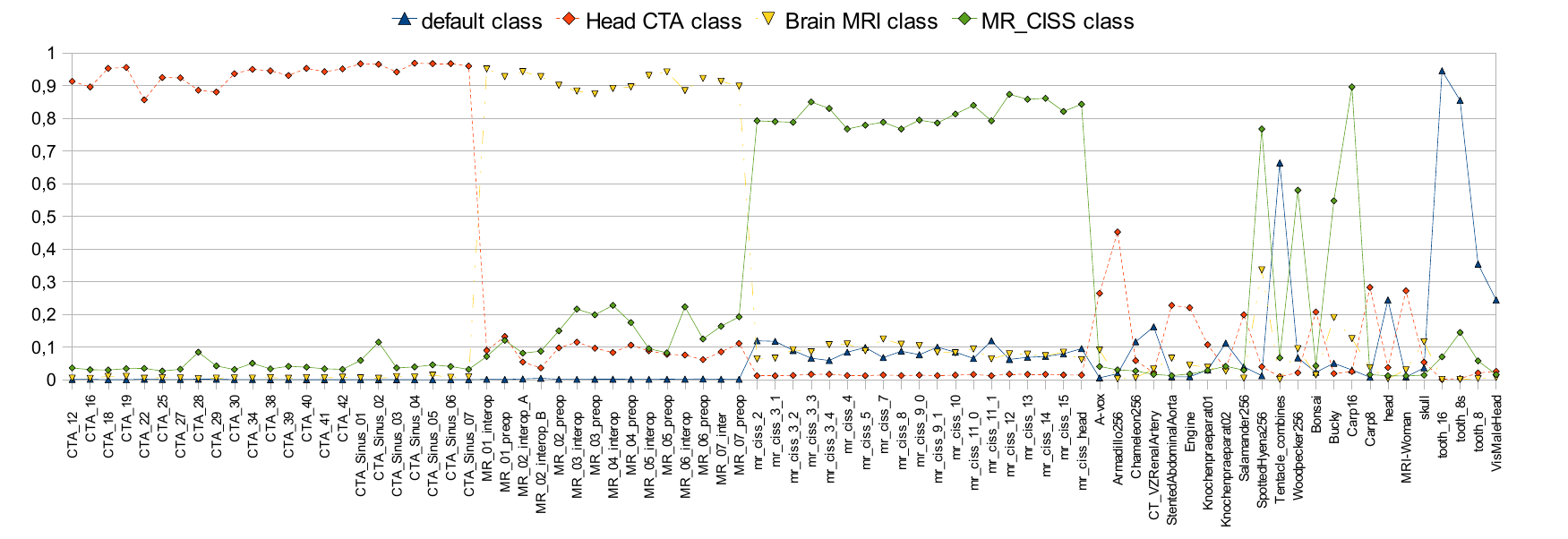}
  \caption{\label{fig:raw39}Raw outputs of the network with the rest class
    approach (``default''). Trained with 1 sample per class. All
    charts in this paper have the same order of datasets as shown here.}
\end{figure*}

\begin{figure*}[htbp]
  \centering
  \includegraphics[width=174mm]{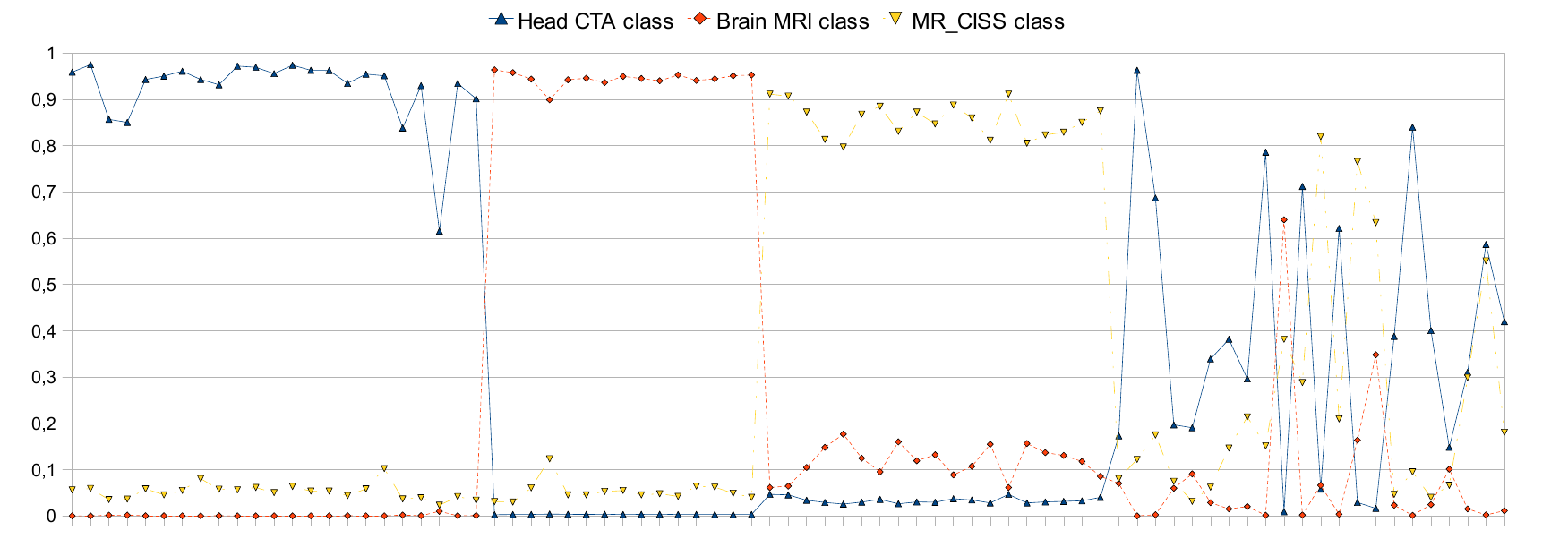}
  \caption{\label{fig:rawnorc1}Raw outputs of the network without the rest class. Trained with 1 sample per class.}
\end{figure*}

\begin{figure*}[htbp]
  \centering
  \includegraphics[width=174mm]{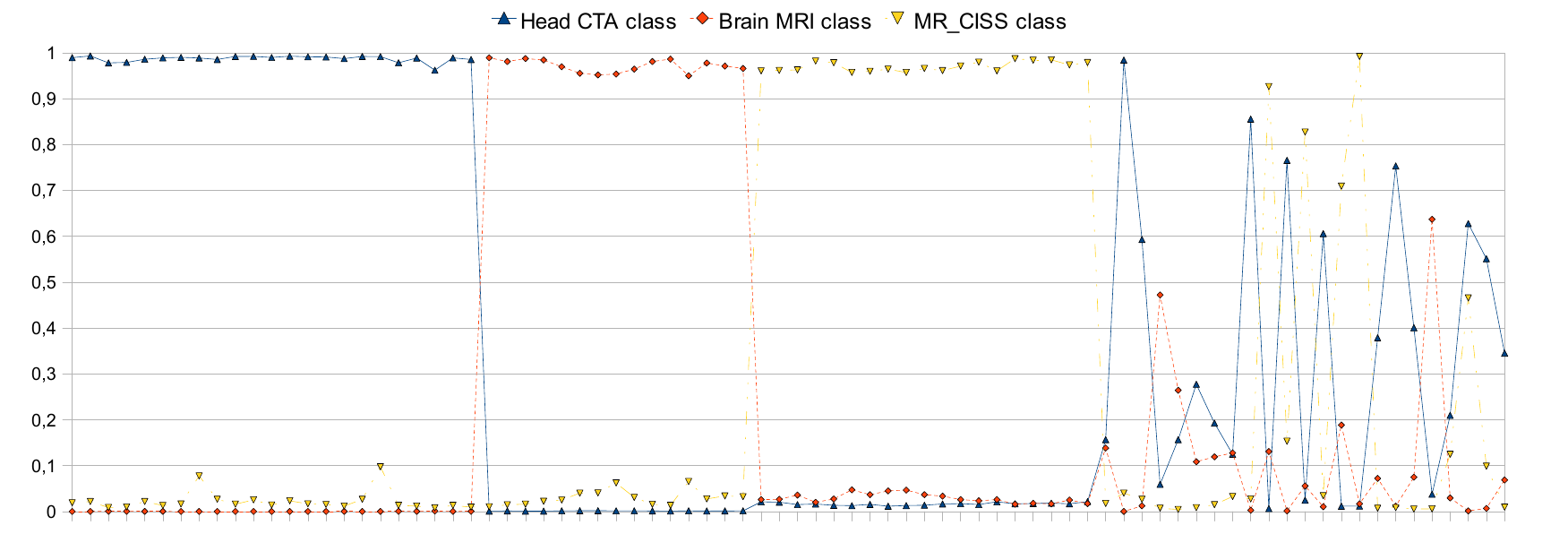}
  \caption{\label{fig:rawnorc2}Raw outputs of the network without the rest class. Trained with 3 samples per class.}
\end{figure*}

In order to identify the most probable classification
result, the output with maximum value is chosen. Therefore, this input would
be classified as belonging to the category one. Fig. \ref{fig:raw39},
\ref{fig:rawnorc1} and \ref{fig:rawnorc2} show actual outputs of a neural network (for easier discerning, descriptive names are given to the outputs).

A training sample consists of the histogram input and the desired output vector. In the desired output vector, only the correct output category has value 1, while all the others have value 0.

\begin{figure}[htb]
  \centering
  \includegraphics[width=6cm]{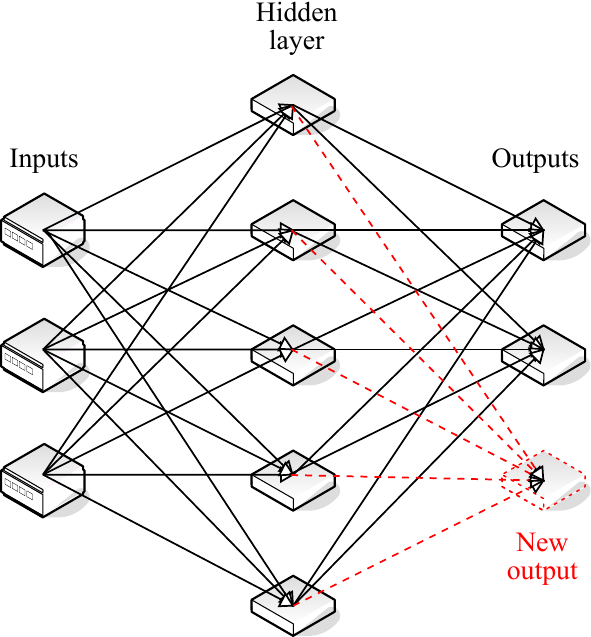}
  \caption{\label{fig:add}Adding an output preserves existing weights. The neural network depicted here is a very small example (compared to real examples) suitable for graphical representation and explanation.}
\end{figure}

In our implementation we chose the multilayer perceptron (MLP), a type of neural network which is capable of performing the required task. It is trained by the back-propagation algorithm. One major benefit of MLP is that additional outputs can be added fairly easily, while retaining the function of all the other outputs. Using some other types of neural networks a new neural network would have to be created and trained from scratch, wasting time whenever a new category is added. Furthermore, this would cause differently randomized initial weights, thus leading to slightly different results. In our version, we only need to add weights between the newly inserted neuron in the output layer and all neurons in the last hidden layer (see Fig.~\ref{fig:add}).

\begin{figure*}[htbp]
  \centering
  \includegraphics[width=174mm]{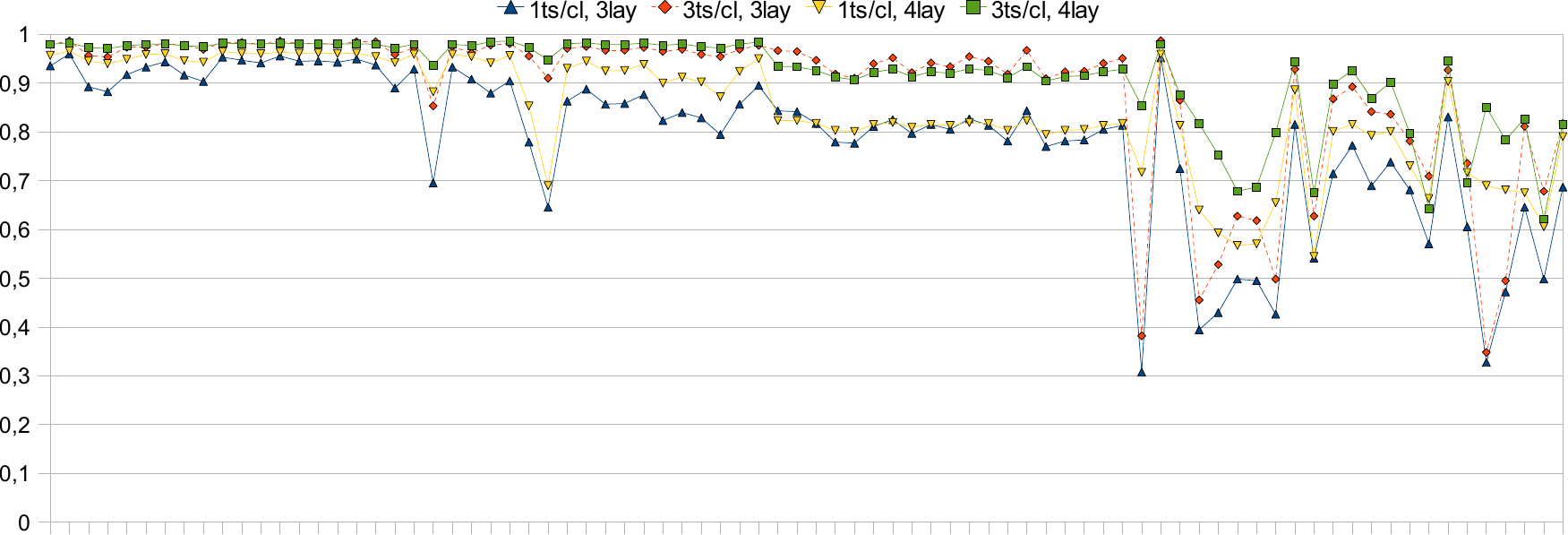}
  \caption{\label{fig:layers}Using 4-layer neural network does not significantly improve results. Only the value of the maximum output is shown for each dataset. Abbr.: ts/cl -- training samples per class.}
\end{figure*}

As feed-forward networks can approximate any continuous real function with as
little as 3 layers, we have only tested networks with 3 and 4 layers. Fewer
number of layers can be compensated with a larger number of neurons in the hidden
layer(s). Although some differences exist (see \cite{Chester90whytwo, nn34}),
they are not relevant for this method (see Fig.~\ref{fig:layers}). All the 
results (except Fig.~\ref{fig:layers}) presented here are obtained using a 
3 layer neural network.

\subsection{Modeling the Rest Class}\label{sec:method.rest}
There are two ways to deal with datasets that do not fall into any of the
well-defined classes, i.e. the miscellaneous datasets. The
first approach is to have a ``rest class'', to which all of these datasets are
associated. The second approach assumes that elements
from the rest class usually do not strongly activate any of the outputs,
often having value of the maximum output around 0,5 (50\%). So the second
approach uses a threshold for successful classification:
If the value of the maximum output is below that threshold, the dataset fails
being classified into any of the well-defined classes and it is considered to be
part of the rest class.

From a conceptual point of view, the threshold approach is independent from the
rest-class approach, i.e. each of the concepts can be applied
separately. From a practical point of view, both approaches
are not completely independent: the better trained the rest class is, the less
effect thresholding provides. Furthermore, providing a
high amount of training samples to the rest class affects the
reliability\footnote{Reliability in this context is the value of the maximum
output} of the classification of the normal (well-defined) classes. If this is coupled with a high threshold,
a lot of ``false negatives'' emerge (datasets misclassified as belonging to the rest class
instead of a well-defined class). However, applying both approaches is
beneficial for lower amounts of training samples for the rest class.

\subsection {Performance issues}
If we directly use histogram pixels as the network's inputs, we have a large number of inputs, e.g. for a 256*256 histogram we get 64K\footnote{prefixes K and M here mean $2^{10}$ and $2^{20}$} inputs. If the second layer contains 64 neurons, the number of weights between 1st and 2nd layer is 4M. In our implementation, the weights are 32-bit floats, which leads to 16MB just for the storage of the weights between the $1^{st}$ and the $2^{nd}$ layer. The amount of weights between other layers is significantly smaller, due to the much lower number of neurons in these layers.

However, the overall memory consumption is relatively exhaustive. Furthermore, the training gets very slow, and an alternative persistent storage on a hard disk would not be convenient due to slow reading, writing and data transfer.

\begin{figure}[htb]
  \centering
  \includegraphics{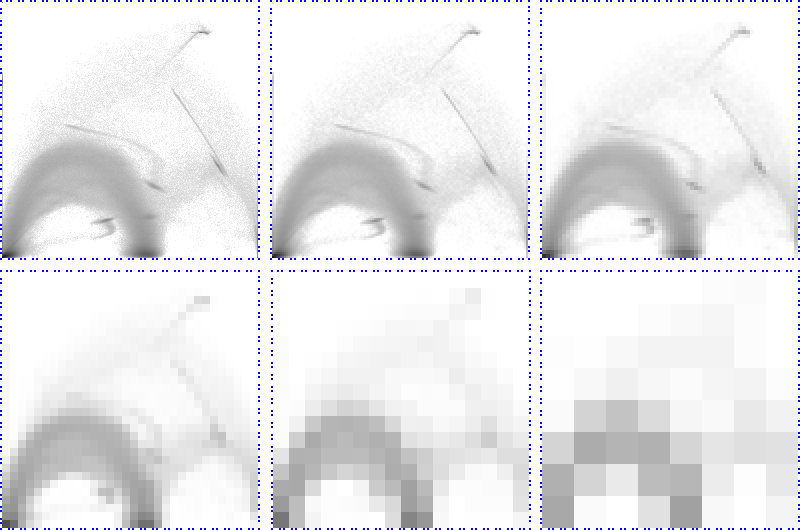}
  \caption{\label{fig:scaling}Size reduction. Upper left is the original 256x256, lower right is 8x8}
\end{figure}

Therefore, we incorporated a downscaling scheme for the histograms by rebinning. This does not only greatly reduce the required data, but it also significantly eliminates small details present in the histograms. For every dataset, their exact positions are always different, so they are only an obstacle for comparison purposes.

For simplicity, our implementation only allows reduction by factors that are powers of 2. That is: 0 -- no  reduction, 1 -- reduction to 128x128, 2 -- reduction to 64x64, etc. Most of the tests have been conducted with reduction factor 3 (histogram size 32x32).

\section{Testing environment}\label{Sec:Testing}

The implementation of the described method is done in a visualization tool called OpenQVis.
OpenQVis focuses on real-time visualization, relying on the features of modern graphics cards~\cite{rtvg}.


OpenQVis has different ``models'' of transfer functions, which are used to visualize different types of 3D datasets. Examples are: CT angiography of the head, MRI scans of the spinal cord, MRI scans of the head, and so on. These models were considered as classes for our method.

OpenQVis allows the user to navigate to a model list and to choose one for the currently opened dataset. If the chosen model is not in the list of the output classes, a new output class is added to the neural network and the network is re-trained with this new training sample. If the chosen class is already present in the outputs, the network is  re-trained with this new training sample included. If the histogram of the currently opened dataset exists among the training samples, the sample is updated to reflect the new user preference.

Saving training samples with the neural network data is required because each re-training consists of many epochs, and if only the newest sample is used the network gradually ``forgets'' previous samples, which is, of course, undesired. So, all saved samples are used for each epoch in the re-training process.

For testing purposes, we had three series available:
\begin{enumerate}
	\item Computed tomography - angiography of the head (CTA\_*), 23 datasets
	\item Magnetic resonance images of the head, both preoperative and inter-operative (MR\_*), 15 datasets
	\item Magnetic resonance - constructive interference in the steady state, mostly scans of the spine (mr\_ciss\_*), 19 datasets
\end{enumerate}
Furthermore, we had 23 miscellaneous datasets (almost all freely available on the Internet). 2 of those datasets were synthetic (bucky and tentacle), generated directly from computer 3D models and not acquired by means of a scanning device.

This method can differentiate between cases within the same scanning modality. We tested this with available but confidential CTA heart datasets, which were clearly discernible from CTA head datasets.

\section{Results}\label{Sec:Results}

The classification based on our neural network approach takes, depending on histogram reduction factor, mere microseconds. The training takes milliseconds for the reduction factor 4 and below. The training for the reduction factor 3 takes noticeable fractions of a second (0,2s to 0,6s) in our tests, and for the reduction factor 2 it takes seconds (3-10 seconds). The training time variations result from the termination condition. We set the condition MSE\footnote{MSE = Mean Squared Error}<0,003 which was nearly almost met before the maximum number of epochs was reached.

\begin{figure*}[htbp]
  \centering
  \includegraphics[width=174mm]{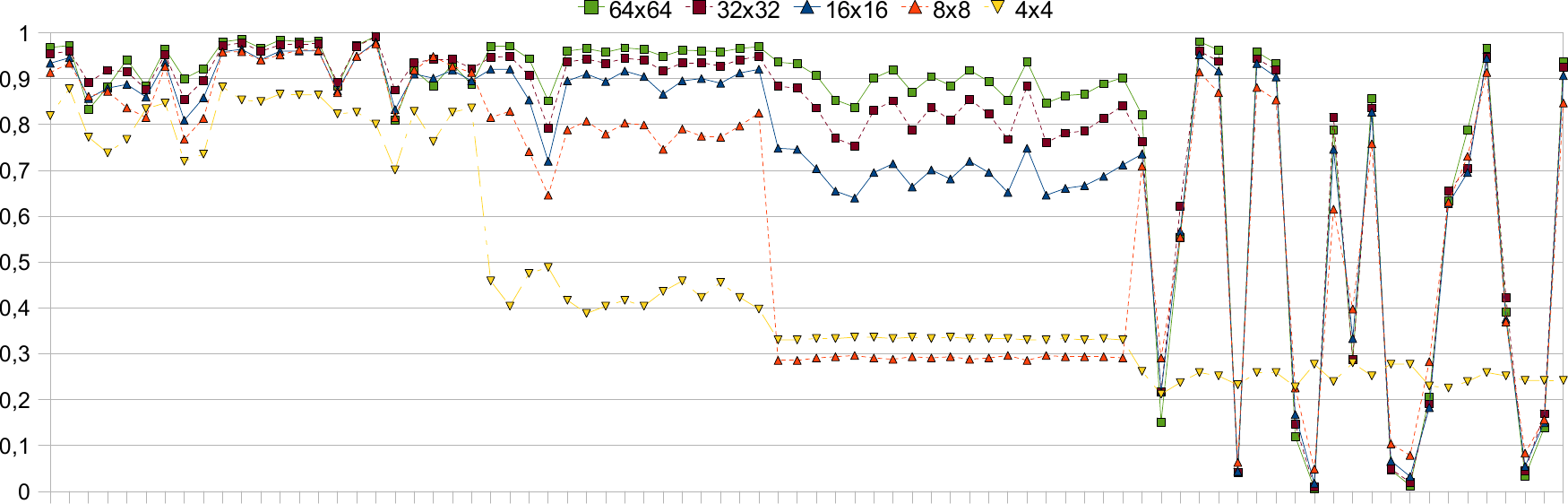}
  \caption{\label{fig:dec}Downscaling the histogram images on the input side of the neural network reduces it's reliability and, in extreme cases, disables the neural network from delineating datasets. Only values of correct outputs are shown -- if desired classification for some dataset is ``default'', value of default output is shown even if it is the lowest-valued output (such case is a misclassification).}
\end{figure*}

The reliability of classification is directly associated with the reduction factor. As seen on Fig.~\ref{fig:dec}, the reliability decreases as the histogram size decreases.

\begin{figure*}[htbp]
  \centering
  \includegraphics[width=174mm]{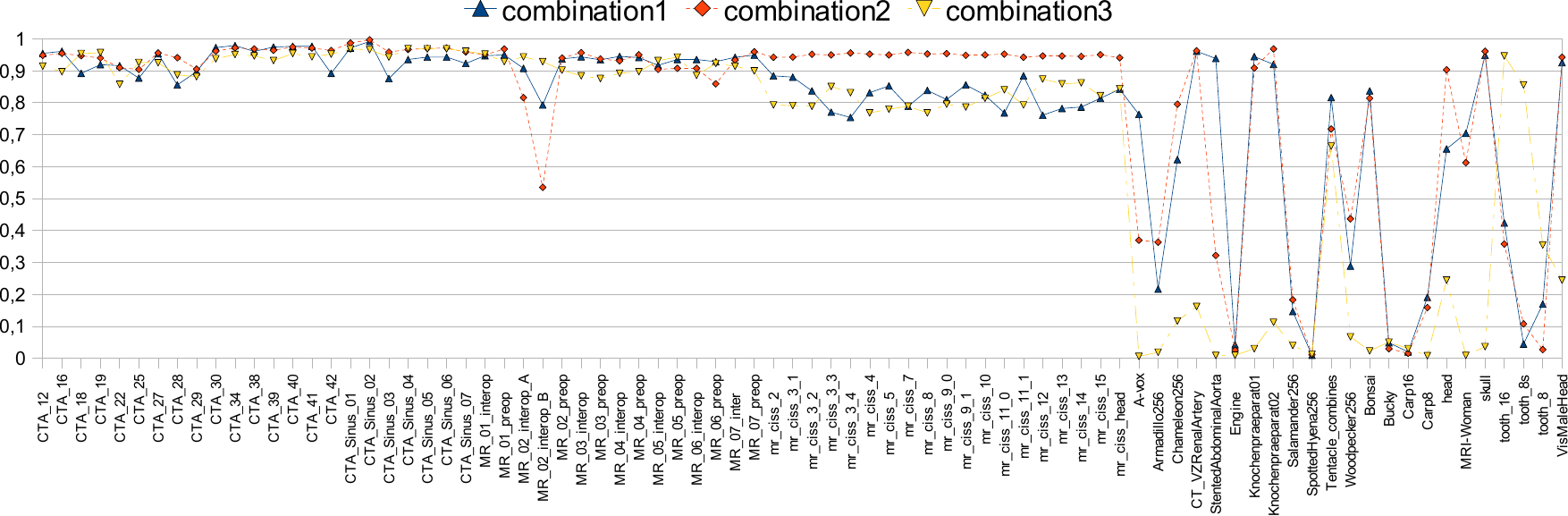}
  \caption{\label{fig:comb}Choosing different datasets for training the neural network influences the results. Trained with 1 sample per class. Only values of correct outputs are shown.}
\end{figure*}

The choice of the dataset which is used to represent a class influences the results to some degree (see Fig.~\ref{fig:comb}). This influence affects the classification outcome only in the miscellaneous group, i.e. the rest class. Choosing an average-looking histogram for the training, or average and extremes in a case of more training samples per class, results in a higher reliability of the classification and in more uniform output values across all datasets of that class.

A slight variation of the results with respect to the initial randomization of the neural network exists, but is negligible. After training the network with one sample of each type, the average difference in outputs (due to different initial weights) is around 1\%. The maximum for any dataset is 5\%. These differences get smaller with a greater number of training samples.

With the rest-class approach, all of the misclassifications occur in the miscellaneous group (see Tab.~\ref{fig:table}). This means, for example, that no CTA is classified as anything else other than CTA. Only datasets from the miscellaneous group are wrongly classified as something else (CTA, MR, or mr\_ciss). This is true even if the neural network is trained with only one sample of each type.

The thresholding approach has a lower amount of misclassifications in the miscellaneous group, but it misclassifies some datasets of the other classes (``false negatives'').

\begin{table}[htb]
  \scriptsize
  \centering
	    \begin{tabular}{|p{4cm}|c|c|}
	    \hline
	    \multirow{2}{*}{Approach/Setup} & \multicolumn{2}{c|}{Misclassification rate} \\
	    & some $\rightarrow$ rest & rest $\rightarrow$ some \\
	    \hline
	    no rest class, no threshold& 0 & all(23) \\ \hline
	    with rest class, 1 ts/cl & 0 & 15-20 \\ \hline
	    with rest class, 2 ts/cl & 0 & 10-15 \\ \hline
	    with rc, 2 ts/wdc and 6 ts/rc & 0 & 3-5 \\ \hline
	    no rest class, threshold 50\% & 0 & 10-15 \\ \hline
	    no rest class, threshold 70\% & 0-1 & 5-10 \\ \hline
	    no rc, threshold 90\%, 1 ts/cl & 20-25 & 1-2 \\ \hline
	    no rc, threshold 90\%, 2 ts/cl & 0 & 3-5 \\ \hline
	    w. rc, threshold 50\%, 1 ts/cl & 0 & 5-15 \\ \hline
	    w. rc, threshold 70\%, 1 ts/cl & 0-5 & 2-10 \\ \hline
	    w. rc, threshold 90\%, 1 ts/cl & 25-30 & 0-2 \\ \hline
	    w. rc, threshold 90\%, 2 ts/cl & 0-5 & 0-2 \\ \hline
	    w. rc, th. 90\%, 2 ts/wdc and 6 ts/rc & 5-10 & 0 \\ \hline
	    \end{tabular}
  \caption{\label{fig:table}Comparison of misclassification rates for different setups of the classifier. If not specified, the classification has been performed using varying parameters in terms of number of training samples per class (for some setups) or choice of datasets used for training, resulting in slightly different misclassification rates. ``rest $\rightarrow$ some'' means that a member of the rest class was wrongly classified as a member of a ``well-defined class''. \newline Abbreviations: w. -- with, ts -- training sample(s), cl -- class, rc -- rest class, wdc -- well-defined class, th. -- threshold.}
\end{table}

From Fig. \ref{fig:raw39}, \ref{fig:rawnorc1}, \ref{fig:rawnorc2} and Tab.~\ref{fig:table} it can be easily concluded that the threshold is a tweaking parameter. Therefore, it should be set high only in specific situations, and in most cases it should be set to a more conservative value (50\%-70\%).

\begin{figure*}[htbp]
  \centering
  \includegraphics[width=174mm]{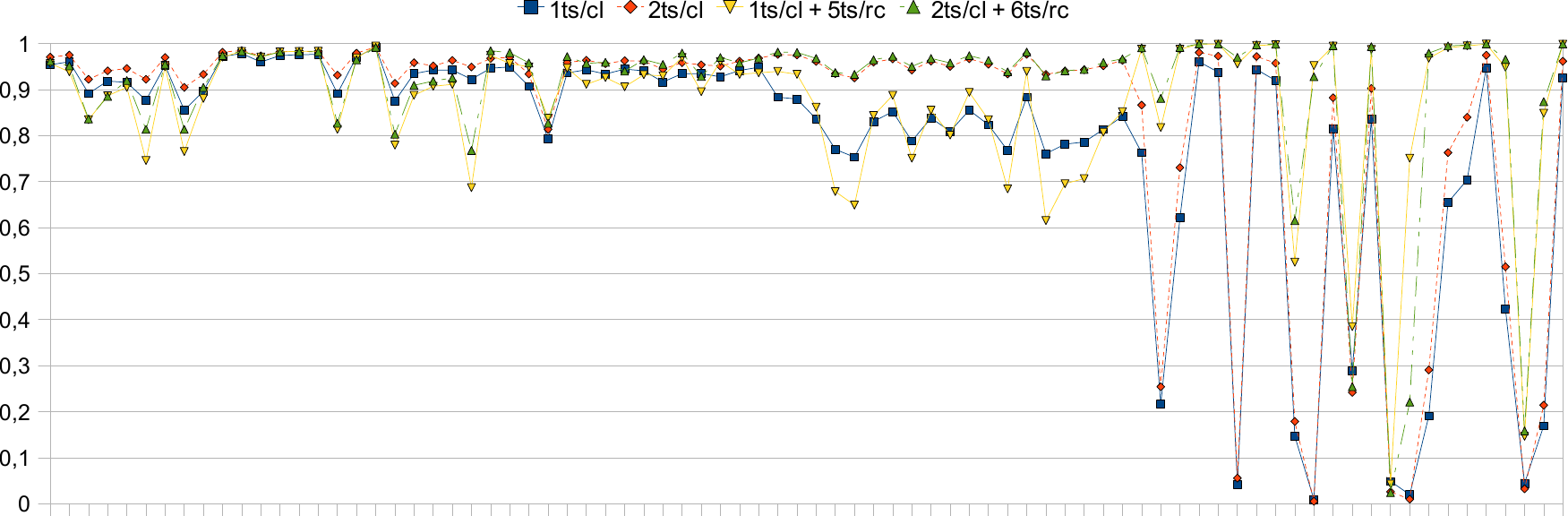}
  \caption{\label{fig:training}Increasing the amount of training samples improves the results. Only values of correct outputs are shown.}
\end{figure*}

Training with multiple datasets of specific classes improves the reliability. Training with multiple datasets of the rest class lowers misclassification rate (see Fig.~\ref{fig:training}).

An alternative method for classifying 3D datasets would be to use a downscaled version of the dataset itself instead of the 2D histogram as input to the neural network. This alternative, however, would strongly incorporate geometric aspects, like the individual orientation of the recorded specimen into the classification process. As a result, the training phase would become more difficult, more training samples would be required, and the number of input nodes will increase considerably to achieve a robustness comparable to the described histogram method.

\section{Conclusion}\label{Sec:Conclusion}
We have presented a robust technique to automatically classify 3D volume datasets according to the acquisition sequence, the recorded specimen and sequence-related parameters. The fact that only one training sample of certain type is sufficient\footnote{with the rest-class approach} to properly classify all the other datasets of the same type is remarkable. Depending in what type of visualization system this method is used, the end user might not need to know anything about it.

Depending on the amount of information about the data and the application scenario, the architecture of the neural network can be adapted to better suite typical use cases (we took a general approach here).

Majority of misclassifications are caused by the datasets belonging to the miscellaneous group. As researchers, we had many different miscellaneous datasets readily available. However, in production systems number of datasets in the rest class should be comparably smaller, thus making this method more appropriate.

An additional advantage of this method is its easy implementation. Successful implementations may be based on one of the many free neural network implementations around (for example, on SourceForge). As a result, the benefits of including this method in a production visualization system (if suitable) will easily outweigh the implementation costs.

{
\bibliographystyle{abbrv}
\bibliography{3IA09Volume}

\begin{thebibliography}{10}

\bibitem{Ankerst993dshape}
M.~Ankerst, G.~Kastenm\"{u}ller, H.-P. Kriegel, and T.~Seidl.
\newblock {3D} shape histograms for similarity search and classification in
  spatial databases.
\newblock In {\em Proc. of the 6th Int. Symposium on Advances in Spatial
  Databases (SSD)}, pages 207--226, London, UK, 1999. Springer-Verlag.

\bibitem{Cerquides06}
J.~Cerquides, M.~López-Sánchez, S.~Ontañón, E.~Puertas, A.~Puig, O.~Pujol,
  and D.~Tost.
\newblock Classification algorithms for biomedical volume datasets.
\newblock In {\em Current Topics in Artificial Intelligence}, volume 4177/2006
  of {\em Lecture Notes in Computer Science}, chapter~16, pages 143--152.
  Springer, 2006.

\bibitem{Chester90whytwo}
D.~L. Chester.
\newblock Why two hidden layers are better than one.
\newblock In {\em Int. Joint Conference on Neural Networks (Washington DC),
  Lawrence Erlbaum Associates}, pages 265--268, Jan 1990.

\bibitem{nn34}
J.~de~Villiers and E.~Barnard.
\newblock Backpropagation neural nets with one and two hidden layers.
\newblock {\em IEEE Trans. on Neural Networks}, 4(1):136--141, Jan 1992.

\bibitem{IPT-EGVE2005:113-120:2005}
A.~del R{\'{i}}o, J.~Fischer, M.~K{\"{o}}bele, D.~Bartz, and W.~Stra{\ss}er.
\newblock {Augmented Reality Interaction for Semiautomatic Volume
  Classification}.
\newblock In E.~Kjems and R.~Blach, editors, {\em {Eurographics Workshop on
  Virtual Environments (EGVE)}}, pages 113--120, Aalborg, Denmark, 2005.
  Eurographics Association.

\bibitem{rtvg}
M.~Hadwiger, J.~M. Kniss, C.~Rezk-Salama, D.~Weiskopf, and K.~Engel.
\newblock {\em Real-time Volume Graphics}.
\newblock A. K. Peters, Ltd., Natick, MA, USA, 2006.

\bibitem{Kindlmann_TransferFunctions}
G.~Kindlmann and J.~W. Durkin.
\newblock Semi-automatic generation of transfer functions for direct volume
  rendering.
\newblock In {\em Proc. of the 1998 IEEE symposium on Volume visualization
  (VVS)}, pages 79--86, New York, NY, USA, 1998. ACM.

\bibitem{kniss-vis01}
J.~Kniss, G.~Kindlmann, and C.~Hansen.
\newblock {Interactive Volume Rendering using Multi-dimensional Transfer
  Functions and Direct Manipulation Widgets}.
\newblock In {\em Proc. of IEEE Visualization (VIS)}, pages 255--262, 2001.

\bibitem{imageret}
Y.~Liu and F.~Dellaert.
\newblock A classification based similarity metric for {3D} image retrieval.
\newblock {\em IEEE Conf. on Computer Vision and Pattern Recognition (CVPR)},
  0:800--805, June 1998.

\bibitem{Lundstrom_LocalHist}
C.~Lundstr{\"{o}}m, P.~Ljung, and A.~Ynnerman.
\newblock {Extending and simplifying Transfer Function design in medical Volume
  Rendering using local histograms}.
\newblock In {\em Eurographics / IEEE VGTC Symposium on Visualization
  (EuroVis)}, pages 263--270, June 2005.

\bibitem{Rautek-2007-SLI}
P.~Rautek, S.~Bruckner, and M.~E. Gr{\"o}ller.
\newblock Semantic layers for illustrative volume rendering.
\newblock {\em IEEE Trans. on Visualization and Computer Graphics},
  13(6):1336--1343, 2007.

\bibitem{Rezk06}
C.~Rezk-Salama, M.~Keller, and P.~Kohlmann.
\newblock {High-Level User Interfaces for Transfer Function Design with
  Semantics}.
\newblock {\em IEEE Trans. on Visualization and Computer Graphics (Proc. IEEE
  Visualization)}, 11(5):1021--1028, 2006.

\bibitem{matfrac}
I.~W. Serlie, F.~M. Vos, R.~Truyen, F.~H. Post, and L.~J. van Vliet.
\newblock Classifying ct image data into material fractions by a scale and
  rotation invariant edge model.
\newblock {\em IEEE Trans. on Image Processing}, 16(12):2891--2904, Dec. 2007.

\bibitem{intronn}
D.~Svozil, V.~Kvasni{\v c}ka, and J.~Posp{\'i}chal.
\newblock Introduction to multi-layer feed-forward neural networks.
\newblock {\em Chemometrics and Intelligent Laboratory Systems}, 39:43--62(20),
  November 1997.

\bibitem{vis03-TFNeuralNetwork}
F.-Y. Tzeng, E.~B. Lum, and K.-L. Ma.
\newblock {A Novel Interface for Higher-Dimensional Classification of Volume
  Data}.
\newblock In {\em Proc. of IEEE Visualization (VIS)}, pages 505--512, 2003.

\bibitem{SeredaBoundaries}
P.~\v{S}ereda, A.~{Vilanova~Bartol\'i}, I.~W.~O. Serlie, and F.~A. Gerritsen.
\newblock {Visualization of Boundaries in Volumetric Data Sets Using LH
  Histograms}.
\newblock {\em Trans. on Vis. and Comp. Graph.}, 12(2):208--218, 2006.

\bibitem{MRI:Zhang}
J.~Zhang and J.~Sun.
\newblock Automatic classification of {MRI} images for three-dimensional volume
  reconstruction by using general regression neural networks.
\newblock In {\em IEEE Nuclear Science Symposium Conference Record}, volume~5,
  pages 3188--3189, Oct. 2003.

\bibitem{nnpos}
{\relax Dž}.~Zukić, A.~Elsner, Z.~Avdagić, and G.~Domik.
\newblock Neural networks in {3D} medical scan visualization.
\newblock In D.~Plemenos, editor, {\em In Proc. of the Int. Conf. on Computer
  Graphics and Artificial Intelligence (3IA)}, pages 183--190. TEI Athens, May
  2008.

\end{thebibliography}
}
\end{document}